\documentclass[12pt]{article}
\usepackage{amsmath}

\title{Grassmann dynamics of classical spin \\ in nonabelian gauge fields}
\author{S. A. Pol'shin
\\
{\small Institute for Theoretical Physics} \\
{\small NSC Kharkov Institute of Physics and Technology} \\
{\small Akademicheskaia St. 1, 61108 Kharkov, Ukraine }}
\date{}

\begin{document}

\maketitle

\begin{abstract}
 Using Grassmann variant of classical mechanics, we construct Lagrangian dynamics of classical spinning particle in (possibly non-abelian) gauge fields. Quantization of this model is briefly discussed.
\end{abstract}

In~\cite{Rav} the Lagrangian theory of pseudoclassical particle moving in electromagnetic field was constructed and its quantization leading to Dirac equation was considered. In the present note, we generalize this theory onto the case of arbitrary gauge group. Quantization and the case of nonzero anomalous magnetic momentum are discussed in last two paragraphs.

Let $x^\mu$ be even "space-time" coordinated of our particle, $\xi^\mu$ odd spin variables,  $A^\mu_a$  even gauge potential, and $J_a$ genetators of certain finite-dimensional anti-hermitean representation of our gauge group (e.g. spinor representation of $SO(3)$), so $[ J_a,J_b ]=i f_{ab}^{\ \ c} J_c$. Let $q,\bar{q}$ be even coordinates of internal gauge degrees of freedom of our particle (their indexes will not be written explicitly). Define their covariant derivatives as
\begin{equation}\label{eq1}
\begin{split}
  \frac{\mathcal{D}q}{\mathcal{D}s}=\frac{dq}{ds}+i e A_\mu^a \dot{x}\vphantom{x}^\mu J_a q \\
  \frac{\mathcal{D}\bar{q}}{\mathcal{D}s}=\frac{d\bar{q}}{ds}-ie A_\mu^a \dot{x}\vphantom{x}^\mu \bar{q} J_a
\end{split}
\end{equation}
where $s$ is even coordinate of a worldline and overdot means derivative w.r.t. $s$. Note that $\dot{x}\vphantom{x}^\mu \dot{x}_\mu \not= 1$ in general, see~\cite{MPLA} for discussion. Define gauge charge as $Q_a =\bar{q} J_a q$, so
\begin{equation}\label{eq2}
     \frac{\mathcal{D} Q_a}{\mathcal{D}s}=\frac{dQ_a}{ds}+ e A_\mu^b \dot{x}\vphantom{x}^\mu f_{ab}^{\ \ c} Q_c.
\end{equation}
Consider the following lagrangian
\begin{equation}\label{eq3}
L=\frac{m}{2}\dot{x}\vphantom{x}^\mu \dot{x}_\mu
-\frac{1}{4}\xi^\mu \dot{\xi}_\mu +\frac{i}{2} \left(
-\frac{\mathcal{D}\bar{q}}{\mathcal{D}s}q+ \bar{q} \frac{\mathcal{D}q}{\mathcal{D}s}\right)+
\frac{\mu'}{2m}F_{\mu\nu}^a S^{\mu\nu} Q_a,
\end{equation}
where $\mu'=e$ is the magnetic moment of a particle (see last paragraph however), $S_{\mu\nu}=\frac{1}{2}\xi_\mu \xi_\nu$ is spin tensor and
\begin{equation}\label{eq4}
    F_{\mu\nu}^a =\partial_\mu A^a_\nu-\partial_\nu A^a_\mu +e A^b_\mu A^c_\nu f_{bc}^{\ \ a}
\end{equation}
is gauge field tensor. Note that unfolding the brackets in r.h.s. of~(\ref{eq3}) we obtain usual interaction term $-e A^a_\mu \dot{x}\vphantom{x}^\mu Q_a$ due to~(\ref{eq1}). Varying the action $\int L\, ds$ w.r.t. $q$ and $\bar{q}$ we obtain
\begin{equation*}
\begin{split}
  i\dot{q} =\left(-e A^a_\mu \dot{x}\vphantom{x}^\mu + \frac{\mu'}{2m}F_{\mu\nu}^a S^{\mu\nu}\right) J_a q\\
   -i\dot{\bar{q}} =\left(-e A^a_\mu \dot{x}\vphantom{x}^\mu + \frac{\mu'}{2m}F_{\mu\nu}^a S^{\mu\nu}\right) \bar{q} J_a ,
\end{split}
\end{equation*}
so using~(\ref{eq2}) we obtain equations of motion of gauge charge
\begin{equation}\label{eq5}
    \frac{\mathcal{D}Q_a}{\mathcal{D}s}=\frac{\mu'}{2m}F_{\mu\nu}^b S^{\mu\nu}f_{ab}^{\ \ c} Q_c.
\end{equation}
Varying w.r.t. $\xi^\mu$ we obtain spin equations of motion
\begin{equation}\label{eq6}
    m\dot{S}\vphantom{S}^{\mu\nu}=\mu' F^{a\rho[\nu}
S_{\: \rho}^{\: \ \mu]} Q_a.
\end{equation}
Finally, varying w.r.t. $x^\mu$ we has to consider $Q_a$ as \textbf{geodesically constant}: $\mathcal{D}Q_a /\mathcal{D}s =0$ but not $\dot{Q}_a =0$ \textbf{contrary to ordinary variational calculus}. Then using~(\ref{eq2}),(\ref{eq4}) we obtain
\begin{equation}\label{eq7}
 m\ddot{x}\vphantom{x}^\mu=eF^{a\mu\nu}\dot{x}_\nu Q_a +\frac{\mu'}{2m}(\mathcal{D}^\mu F^{\rho\sigma})^a S_{\rho\sigma} Q_a,
\end{equation}
where $\mathcal{D}^\mu$ is ordinary covariant derivative w.r.t. $A_\mu^a$, so $d(F^a_{\mu\nu} Q_a)/ds=(\mathcal{D}_\rho F_{\mu\nu})^a Q_a \dot{x}\vphantom{x}^\rho +F^a_{\mu\nu}\mathcal{D}Q_a /\mathcal{D}s$. Eqs.~(\ref{eq5})-(\ref{eq7}) are just the ones obtained by Heinz~\cite{Heinz} by classicalizing the ordinary QCD hamiltonian (cf. also~\cite{Holt}).

Due to nonabelian Bianchi identity
$$(\mathcal{D}^{[\mu} F^{\rho\sigma]})^a =0$$
we see that the following quantities are conserved
\begin{equation}
    \begin{split}
      C_1 =\xi_\mu \dot{x}\vphantom{x}^\mu \\
      C_2 =\frac{m}{2}\dot{x}\vphantom{x}^\mu \dot{x}_\mu + \frac{\mu'}{2m}F_{\mu\nu}^a S^{\mu\nu} Q_a \\
      C_3 =\bar{q}q.
    \end{split}
\end{equation}
Since orbits of coadjoint representation are distinguished by $C_3$ values, we see that different types of gauge charges arise in the sense of~\cite{Wein}.

To quantize our theory, we turn $q,\bar{q}$ into bosonic creation-destruction operators, then different values of occupation number operator $\hat{C}_3$ correspond to different representations of gauge group, $\xi^\mu$ turn into $\gamma$-matrices, and canonical momentum $p_\mu=\eta_{\mu\nu}\dot{x}\vphantom{x}^\nu -eA_\mu^a Q_a$~\cite{Mont} turns into $-i\partial_\mu$. Thus $\hat{C}_1$ becomes Dirac operator (cf.~\cite{Rav} in the abelian case) and $\hat{C}_2$ becomes Hamiltonian. If coadjoint representation of gauge group is chosen, then $\hat{Q}_a$ turn into Gell-Mann matrices of ordinary QCD.

If $\mu' \not=e$, we obtain theory with anomalous magnetic momentum. Considering $C_1$ as a Lagrangian constraint, we see that all the above considerations go without substantial change, so we obtain BMT~\cite{BMT}-type equations (cf.~\cite{MPLA} for abelian case)
$$m (\dot{x}\vphantom{x}^\rho \dot{x}_\rho ) \dot{S}\vphantom{S}^{\mu\nu}=\mu' (\dot{x}\vphantom{x}^\kappa \dot{x}_\kappa ) F^{a\rho[\nu}
S_{\: \rho}^{\: \ \mu]} Q_a +(\mu'-e)
F^{a\rho\sigma} \dot{x}\vphantom{x}_{\: \rho}  S_{\: \sigma}^{\: \ [\mu} \dot{x}\vphantom{x}^{\nu]} Q_a,$$
and some additional terms in the r.h.s. of eq.~(\ref{eq7}) arise.
For the case of $U(1)$ gauge group, the quantized version of $C_2$ for an arbitrary value of $\mu'$ was considered in~\cite{Andr}.

\end{document}